\documentclass[aps,pra,twocolumn,a4paper,showpacs,superscriptaddress,floatfix,10pt]{revtex4}

\usepackage{amsmath,amssymb}
\usepackage{graphicx}
\usepackage{float}
\usepackage{subfigure}
\usepackage{verbatim}
\usepackage{amsfonts}
\usepackage{dcolumn}
\usepackage{bm}
\usepackage{color}
\usepackage[colorlinks,citecolor=blue]{hyperref}
\usepackage{mathrsfs}
\newcommand{\bra}[1]{\mbox{$\langle #1 |$}}
\newcommand{\ket}[1]{\mbox{$| #1 \rangle$}}
\newcommand{\tr}{\mbox{tr}}
\newcommand{\braket}[2]{\mbox{$\langle #1 | #2 \rangle$}}
\begin{document}

\hyphenpenalty=5000
\tolerance=1000

\title{Disentangling Theorem and Monogamy for Entanglement Negativity}

\author{Huan He}
\affiliation{Perimeter Institute for Theoretical Physics, Waterloo, Ontario, N2L 2Y5, Canada}
\author{Guifre Vidal}
\affiliation{Perimeter Institute for Theoretical Physics, Waterloo, Ontario, N2L 2Y5, Canada}

\begin{abstract}
Entanglement negativity is a measure of mixed-state entanglement increasingly used to investigate and characterize emerging quantum many-body phenomena, including quantum criticality and topological order. We present two results for the entanglement negativity: a disentangling theorem, which allows the use of this entanglement measure as a means to detect whether a wave-function of three subsystems $A$, $B$, and $C$ factorizes into a product state for parts $AB_1$ and $B_2C$; and a monogamy relation, which states that if $A$ is very entangled with $B$, then $A$ cannot be simultaneaously very entangled also with $C$.
\end{abstract}

\maketitle

\section{Introduction}

In recent years, the study of quantum entanglement has provided us with novel perspectives and tools to address many-body problems. Progress in our understanding of many-body entanglement has resulted both in the development of efficient tensor network descriptions of many-body wavefunctions \cite{TN} and in the identification of diagnoses for quantum criticality \cite{QCrit} and topological order \cite{TO}.

Measures of entanglement have played a key role in the above accomplishments. The most popular measure is the entanglement entropy \cite{book}, namely the von Neuman entropy $S(\rho_A)$ of the reduced density matrix $\rho_A$ of region $A$, 
\begin{equation}
	S(\rho_A) \equiv -\tr(\rho_A \log_2 (\rho_A)), ~~~ \rho_A \equiv \tr_{B} \ket{\Psi_{AB}}\bra{\Psi_{AB}}\
\end{equation}
which is used to quantify the amount of entanglement between region $A$ and its complementary $B$ when the whole system $AB$ is in the pure state $\ket{\Psi_{AB}}$.

In order to go beyond bipartite entanglement for pure states, another measure of entanglement was introduced, namely the entanglement negativity $\mathcal{N}^{A|B}$ \cite{Neg,VW,other}. The entanglement negativity is used to quantify the amount of entanglement between parts $A$ and $B$ when these are in a (possibly mixed) state $\rho_{AB}$. One can always think of regions $A$ and $B$ as being parts of a larger system $ABC$ in a pure state $\ket{\Psi_{ABC}}$ such that $\rho_{AB} = \tr_C(\ket{\Psi_{ABC}}\bra{\Psi_{ABC}})$, and therefore use the entanglement negativity to also characterize tripartite entanglement. 

Recently, Calabrese, Cardy, and Tonni have sparked renewed interest in the entanglement negativity through a remarkable exact calculation of its scaling in conformal field theory \cite{NegCFT}. Similarly, an exact calculation is possible for certain systems with topological order \cite{NegTO}. Moreover, the negativity is easily accessible from tensor network representations \cite{NegTN} and through quantum monte carlo calculations \cite{NegMC}.

\begin{figure}
  \includegraphics[width=0.4\textwidth]{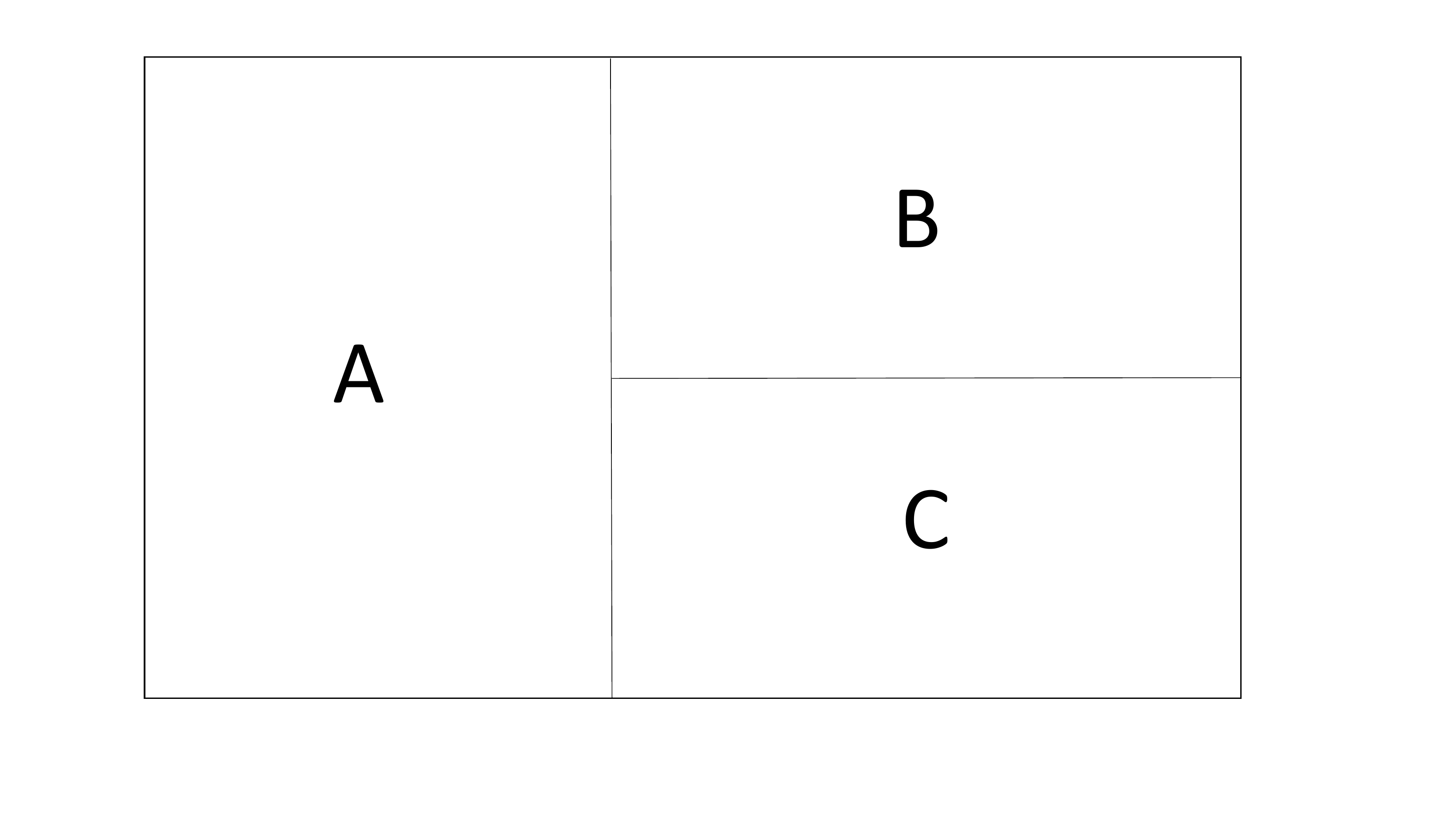}\\
  \caption{Disentangling Theorem: For a tripartite system ABC, if $\mathcal{N}^{A|BC}=\mathcal{N}^{A|B}$, then there exist a factorization of $\mathcal{H}_B \cong (\mathcal{H}_{B_1}\bigotimes\mathcal{H}_{B_2})\oplus \mathcal{H}_{B}^{\perp}$ such that the whole state $|\psi_{ABC}\rangle$ can be factorized as $\ket{\Psi_{AB_1}}\otimes \ket{\psi_{B_2C}}$.}
  \label{fig:ABC}
\end{figure}

In this paper we add to the above recent contributions by presenting a disentangling theorem for the entanglement negativity. This technical theorem allows us to use the negativity to learn about the structure of a many-body wave-function. The theorem states that if and only if the negativity $\mathcal{N}^{A|BC}$ between parts $A$ and $BC$ of a system $ABC$ in a pure state $\ket{\Psi_{ABC}}$ does not decrease when we trace out part $C$, that is, if and only if $\mathcal{N}^{A|BC} = \mathcal{N}^{A|B}$, see fig. \ref{fig:ABC}, then it is possible to factorize the vector space $\mathcal{H}_B$ of part $B$ as $\mathcal{H}_{B_1} \otimes \mathcal{H}_{B_2}$ (direct sum with an irrelevant subspace) in such a way that the state $\ket{\Psi}$ itself factorizes as 
\begin{equation}
	\ket{\Psi_{ABC}} = \ket{\Psi_{A B_1}}\otimes \ket{\Psi_{B_2 C}}. \label{eq:factorization}
\end{equation}
This is a remarkable result. Notice that one can come up with many different measures of entanglement for mixed states (most of which may be very hard to compute). These measures of entanglement will in general differ from each other and from the negativity for particular states, and it is often hard to attach a physical meaning to the concrete value an entanglement measure takes. [For instance, in the case of the logarithmic negativity, we only know that it is an upper bound to how much pure-state entanglement one can distill from the mixed state \cite{VW}]. However, the disentangling theorem tells us that through a calculation of the entanglement negativity we can learn whether the wave-function factorizes as in Eq. \ref{eq:factorization}. That is, we are not just able to use the entanglement negativity to attach a number to the amount of mixed-state entanglement, but we are also able to learn about the intrincate structure of the many-body wave-function. In some sense, this theorem is analogous to Hayden et al.'s necessary and sufficient conditions for the saturation of the strong subadditivity inequality for the von Neumann entropy \cite{SubAdditivity}, a beautiful result with deep implications in quantum information theory. 


We also present a numerical study of monogamy that puts 
the above disentangling theorem in a broader perspective. It follows from Eq. \ref{eq:factorization} that state $\ket{\Psi_{ABC}}$ has no entanglement between parts $A$ and $C$, so that $\mathcal{N}^{A|C}=0$. That is, the disentangling theorem refers to a setting where the entanglement negativity fulfills the monogamy relation,
\begin{equation}
	\mathcal{N}^{A|BC} \geq \mathcal{N}^{A|B} + \mathcal{N}^{A|C}, \label{eq:monogamy}
\end{equation}
(for the particular case $N^{A|C} = 0$). We have seen numerically that the entanglement negativity does not fulfill the monogamy condition of Eq. \ref{eq:monogamy}. However, we have also found that the square of the negativity entanglement negativity satisfies the monogamy relation:
\begin{equation}
	(\mathcal{N}^{A|BC})^2 \geq (\mathcal{N}^{A|B})^2 + (\mathcal{N}^{A|C})^2. \label{eq:monogamy2}
\end{equation}
For the particular case of a three-qubit system, this result had been previously proved analytically by Ou and Fan \cite{OuFan}.

The rest of the paper is divided into sections as follows. 
First in Sect. II we review the entanglement negativity. 
Then in Sect. III we present and prove the disentangling theorem and discuss two simple corollaries. 
Finally, in Sect. IV we analyze a monogamy relation for entanglement negativity, and Sect. V contains our conclusions.

\section{Brief Review on Entanglement Negativity}

The negativity (now known as entanglement negativity) was first introduced in Ref. \cite{Neg} and later shown in Ref. \cite{VW} to be an entanglement monotone -- and therefore a suitable candidate to quantify entanglement (see also Refs. \cite{other}). The entanglement negativity $\mathcal{N}^{A|B}$ of $\rho_{AB}$ is defined as the absolute value of the sum of negative eigenvalues of $\rho^{T_A}_{AB}$, where the symbol $T_A$ means partial transpose with respect to subsystem $A$. Equivalently, 
\begin{equation}
\mathcal{N}^{A|B} \equiv \frac{\|\rho^{T_A}_{AB}\|_1-1}{2}
\end{equation}
where $\|X\|_1\equiv \tr\sqrt{X^\dag X}$. A related quantity, the logarithmic negativity $E_N\equiv \log_2(|\rho^{T_A}_{AB}\|_1) =\log_2(1+2\mathcal{N}^{A|B})$ \cite{VW}, is an upper bound to the amount of pure state entanglement that can be distilled from $\rho_{AB}$. 


Ref. \cite{VW} contains a long list of properties of the entanglement negativity. In this paper, we will need the following two results (see Lemma 2 in Ref. \cite{VW}):

\textbf{\emph{Lemma 1}}: For any Hermitian matrix A there is a decomposition of the form $A=a_{+}\rho^{+}-a_{-}\rho^{-}$, where $\rho^{\pm}\geqslant0$ are density matrices and $a_{\pm}\geqslant0$.

We say that a specific decomposition of the form $A=a_{+}\rho^{+}-a_{-}\rho^{-}$ is optimal if $a_+ + a_-$ is minimal over all possible decompositions of the same form.

\textbf{\emph{Lemma 2}}: The following four statements of a decomposition of the form in Lemma 1 are equivalent:
\begin{enumerate}
	\item Decomposition $A=a_{+}\rho^{+}-a_{-}\rho^{-}$ is optimal (that is,  $a_+ + a_-$ is minimal).
	\item $\|A\|_1=a_++a_-$.
	\item $\mathcal{N} = a_-$ (that is, $a_-$ is the absolute value of the sum of negative eigenvalues of $A$, or its negativity).
	\item $\rho^+$ and $\rho^-$ have orthogonal support, so that $\tr(\rho^+ \rho^-) = 0$ (we say $\rho^+$ and $\rho^-$ are orthogonal).
\end{enumerate}

%


\section{Disentangling Theorem}

Let $\ket{\Psi_{ABC}}$ be a pure state of a system $ABC$ made of three parts $A$, $B$, and $C$, with vector space $\mathcal{H}_{ABC} \cong \mathcal{H}_A \otimes \mathcal{H}_B \otimes \mathcal{H}_C$. Let $\rho_{AB} \equiv \tr_{C}(\ket{\Psi_{ABC}}\bra{\Psi_{ABC}})$ be the reduced density matrix for parts $AB$, and let $\mathcal{N}^{A|BC}$ and $\mathcal{N}^{A|B}$ be the entanglement negativity between $A$ and $BC$ for state $\ket{\Psi_{ABC}}$, and between parts $A$ and $B$ for state $\rho_{AB}$, respectively, see Fig.\ref{fig:ABC}.

\textbf{\emph{Theorem 3 (Disentangling Theorem):}}
The entanglement negativities $\mathcal{N}^{A|BC}$ and $\mathcal{N}^{A|B}$ are equal if and only if there exists a decomposition of $\mathcal{H}_B$ as $(\mathcal{H}_{B_1} \otimes \mathcal{H}_{B_2}) \oplus \mathcal{H}_{B}^{\perp}$, such that the state $\ket{\Psi_{ABC}}$ decomposes as $\ket{\Psi_{A B_1}} \otimes \ket{\Psi_{B_2 C}}$. That is,
\begin{eqnarray}
\mathcal{N}^{A|BC} &=&\mathcal{N}^{A|B} ~\Leftrightarrow ~ \exists ~ \mathcal{H}_{B_1}, \mathcal{H}_{B_2}, \mathcal{H}_{B}^{\perp}  \label{eq:disentangling}\\ 
&& \mbox{such that}  \left\{
\begin{array}{l}
\mathcal{H}_B \cong (\mathcal{H}_{B_1} \otimes \mathcal{H}_{B_2}) \oplus \mathcal{H}_{B}^{\perp}  \\
\ket{\Psi_{ABC}} = \ket{\Psi_{A B_1}} \otimes \ket{\Psi_{B_2 C}}. 
\end{array}\right.\nonumber
\end{eqnarray}

\emph{Proof:}

Proving one direction, namely that if $\ket{\Psi_{ABC}} = \ket{\Psi_{A B_1}} \otimes \ket{\Psi_{B_2 C}}$, then  $\mathcal{N}^{A|BC} =\mathcal{N}^{A|B}$, is simple: one just needs to explicitly compute the negativities $\mathcal{N}^{A|BC}$ and $\mathcal{N}^{A|B}$ following their definition. We thus focus on proving the opposite direction. For the sake of clearness, the proof is divided into three steps: (a) Preparation; (b) Orthogonal condition; (c) Factorization.

\emph{Step (a). Preparation:}

Let us start by writing $\ket{\Psi_{ABC}}$ in each Schmidt decomposition according to the bipartition $A|BC$, 
\begin{equation}
	\ket{\Psi_{ABC}} = \sum_{i} \sqrt{p_i} \ket{\phi_A^i \phi_{BC}^i},
\end{equation}
where $p_i > 0$, $\sum_i p_i=1$, $\braket{\phi_A^i}{\phi_A^j} = \delta_{ij}$, and $\braket{\phi_{BC}^i}{\phi_{BC}^j} = \delta_{ij}$.
Further, each state $\ket{\phi_{BC}^i}$ can be decomposed using an orthonormal set of states $\{\ket{\phi_B^j}\}$ and $\{\ket{\phi_C^k}\}$ in parts $B$ and $C$ respectively, as
\begin{equation}
	\ket{\phi_{BC}^i} = \sum_{jk} T^i_{jk} \ket{\phi_B^j \phi_C^k},
\end{equation}
where the coefficients $T^i_{jk}$ fulfill the unitary constraints:
\begin{equation} \label{eq:unitary}
 \sum_{ab}{T^{i*}_{ab}T^{j}_{ab}}=\braket{\phi_{BC}^i}{\phi_{BC}^j} = \delta_{ij}.
\end{equation}
Let $\rho_{ABC} \equiv \ket{\Psi_{ABC}}\bra{\Psi_{ABC}}$. We note that $\mathcal{H}_B$ decomposes as the direct sum of two subspaces, $\mathcal{H}_B \cong \mathcal{H_B}^{\parallel} \otimes \mathcal{H_B}^{\perp}$, where subspace $\mathcal{H_B}^{\parallel}$ contains the support of $\rho_B \equiv \tr_{AC} (\rho_{ABC})$ and subspace $\mathcal{H_B}^{\perp}$ is its orthogonal complement. In particular, states $\{\ket{\phi_B^j}\}$ above form an orthonormal basis in $\mathcal{H_B}^{\parallel}$. Similar decompositions of course also apply to $\mathcal{H}_A$ and $\mathcal{H}_C$, but we will not need them here.

The operators $\rho_{ABC}$ and $\rho^{T_A}_{ABC}$ can be expressed as:
\begin{eqnarray}
\rho_{ABC}       &=& \sum_{ij} \sqrt{p_ip_j}\ket{\phi_A^i \phi_{BC}^i}\bra{\phi_A^j \phi_{BC}^j},   \\
\rho^{T_A}_{ABC} &=& \sum_{ij} \sqrt{p_ip_j}\ket{\phi_A^j \phi_{BC}^i}\bra{\phi_A^i \phi_{BC}^j}.
\end{eqnarray}

It is then easy to verify that all eigenvalues and corresponding eigenvectors of $\rho^{T_A}_{ABC}$ are given by:

\begin{itemize}
  \item $~p_i,~~~~~~~~\ket{\phi_A^i\phi_{BC}^i}, \forall i$,
  \item $~\sqrt{p_ip_j},~~(\ket{\phi_A^i\phi_{BC}^j}+\ket{\phi_A^j\phi_{BC}^i})/\sqrt{2}, \forall i,j; ~i<j$,
  \item $-\sqrt{p_ip_j},~~(\ket{\phi_A^i\phi_{BC}^j}-\ket{\phi_A^j\phi_{BC}^i})/\sqrt{2}, \forall i,j; ~i<j$.
\end{itemize}

Let us denote the above three types of eigenvectors simply as $\ket{0_{i}}$, $\ket{+_{ij}}$, and $\ket{-_{ij}}$, respectively, so that
\begin{eqnarray}
\rho^{T_A}_{ABC} &=\sum_{i} p_i \ket{0_i} \bra{0_i} + \sum_{i<j} \sqrt{p_ip_j}\ket{+_{ij}}\bra{+_{ij}} \nonumber \\
&-\sum_{i<j} \sqrt{p_ip_j} \ket{-_{ij}}\bra{-_{ij}} \nonumber\\
&= (1+n)\rho^{+}_{ABC} - n \rho^{-}_{ABC},
\end{eqnarray}
where 
\begin{eqnarray}
\rho^{+}_{ABC} &\equiv& \frac{1}{1+n}\left( \sum_{i} p_i \ket{0_i} \bra{0_i} + \sum_{i<j} \sqrt{p_ip_j}\ket{+_{ij}}\bra{+_{ij}} \right), \nonumber \\
\rho^{-}_{ABC} &\equiv& \frac{1}{n}  \sum_{i<j} \sqrt{p_ip_j} \ket{-_{ij}}\bra{-_{ij}},\\
 n &\equiv& \mathcal{N}^{A|BC} = \left|\sum_{i<j} (-\sqrt{p_ip_j})\right|,
\end{eqnarray}
and where $\rho^{+}_{ABC}$ and $\rho^{-}_{ABC}$ are non-negative, $\rho^{\pm}_{ABC} \geq 0$, and orthogonal, $\tr (\rho^{+}_{ABC} \rho^{-}_{ABC})=0$ (see lemma 2).
On the other hand, the matrix $\rho^{T_A}_{AB}\equiv \tr_C(\rho^{T_A}_{ABC})$ reads
\begin{eqnarray}
\rho^{T_A}_{AB} &=& \sum_{i}{p_iM^{0}_{i}}+\sum_{i<j}{\sqrt{p_ip_j}M^{+}_{ij}}-\sum_{i<j}{\sqrt{p_ip_j}{M^{-}_{ij}}}, \nonumber\\
&=& (1+n)\rho^{+}_{AB} - n \rho^{-}_{AB}, \label{eq:newdeco}
\end{eqnarray}
where we have introduced 
\begin{eqnarray}
M^{-}_{ij} &\equiv& \tr_C|-_{ij}\rangle\langle -_{ij}|=\frac{1}{2}\sum_m{|N^m_{ij}\rangle\langle N^m_{ij}|},\\
  M^{+}_{ij} &\equiv& \tr_C|+_{ij}\rangle\langle +_{ij}| = \frac{1}{2}\sum_m{|P^m_{ij}\rangle\langle P^m_{ij}|}, \\
  M^{0}_{i}  &\equiv& \tr_C\ket{0_i}\bra{0_i} = \sum_m{|O^m_i\rangle\langle O^m_i|},
\end{eqnarray}
with 
\begin{eqnarray}
|N^m_{ij}\rangle &=& \sum_a{(T^j_{am}|A^i\rangle-T^i_{am}|A^j\rangle)|B^a\rangle}, \\
|P^m_{ij}\rangle &=& \sum_a{(T^j_{am}|A^i\rangle+T^i_{am}|A^j\rangle)|B^a\rangle},\\
|O^m_i\rangle &=& \sum_a{T^i_{am}|A^i\rangle|B^a\rangle},
\end{eqnarray}
and where
\begin{eqnarray}
\rho^{+}_{AB} &\equiv& \frac{1}{1+n}\left( \sum_{i} p_i M_i^0 + \sum_{i<j} \sqrt{p_ip_j} M^+_{ij} \right), \\
\rho^{-}_{AB} &\equiv& \frac{1}{n}  \sum_{i<j} \sqrt{p_ip_j} M^-_{ij}.
\end{eqnarray}
By construction $\rho^{+}_{AB}$ and $\rho^{-}_{AB}$ are still non-negative, $\rho^{\pm}_{AB} \geq 0$, but they no longer necessarily orthogonal or, in other words, the decomposition in Eq. \ref{eq:newdeco} may not be optimal -- and consequently, the negativity might be smaller than $n$.

\emph{Step (b). Orthogonality condition:}

Lemma 2 above tells us that, in order to preserve the negativity $n=\mathcal{N^{A|BC}}$ of $\ket{\Psi_{ABC}}$ after tracing out part C, the positive operators $\rho^-_{AB}$ and $\rho^+_{AB}$ have to be orthogonal, that is 
\begin{equation}
\tr(\rho^+_{AB} \rho^-_{AB}) = 0. \label{eq:ortho}
\end{equation}
It is not difficult to show that this amounts to requiring that $\langle N^m_{ij}|O^n_s\rangle = \langle N^m_{ij}|P^n_{st}\rangle = 0$ for all valid values of $i,j,m,n,s,t$. Let us analyze these conditions carefully. First, let us consider 
\begin{equation}\label{eq.typeI}
0 = \langle N^m_{ij}|O^n_s\rangle=\sum_a({T^{j*}_{am}T^s_{an}\delta_{is}-T^{i*}_{am}T^s_{an}\delta_{js}}).
\end{equation}
There are four particular cases to be considered:
\begin{itemize}
  \item If $i\neq s$ and $j\neq s$, then Eq. \ref{eq.typeI} is already zero due to the $\delta_{i,s}$ and $\delta_{j,s}$;
  \item If $j=s$ and  $i\neq s$, then we find the condition $\sum_a{T^{i*}_{am}T^s_{an}}=0$;
  \item If $i=s$  and $j\neq s$, then we recover the same condition as for $j=s, i\neq s$;
  \item The case $i=s$ and $j=s$ does not exist because we demanded $i\neq j$ from the start.
\end{itemize}
In summary, we have obtained the condition:
\begin{equation} \label{eq:condI}
	\sum_a{T^{i*}_{am}T^j_{an}}=0, \forall i, j, m, n, ~\mbox{such that}~ i\neq j.
\end{equation}
Second, let us consider 
\begin{eqnarray}\label{eq.typeII}
&&\langle N^m_{ij}|P^n_{st}\rangle=\\
&&\sum_a({T^{j*}_{am}T^{s}_{an}\delta_{it}+T^{j*}_{am}T^{t}_{an}\delta_{is}-T^{i*}_{am}T^{s}_{an}\delta_{jt}-T^{i*}_{am}T^{t}_{an}\delta_{js}}). \nonumber
\end{eqnarray}
There are again four particular cases to be considered:
\begin{itemize}
  \item If $i\neq s, t$ and $j\neq s, t$, then Eq. \ref{eq.typeII} is already zero due to the $\delta_{i,s}$, $\delta_{i,t}$,  $\delta_{j,s}$ and $\delta_{j,t}$;
  \item If $i=s$ (or if $i = t$) and $j\neq s,t$, then we reach again Eq. \ref{eq:condI};
  \item If $i=s,j=t$ or $i=t,j=s$, because of $i\neq j, s\neq t$, then we must have $\sum_a{T^{i*}_{am}T^{i}_{an}}=\sum_a{T^{j*}_{am}T^{j}_{an}}$;
  \item The case $i=s, j=s$ (or $i=t, j=t$) does not exist because we demanded $i\neq j$ from the start.
\end{itemize}
In summary, we have obtained the new condition
\begin{equation} \label{eq:condII}
	\sum_a{T^{i*}_{am}T^{i}_{an}}=\sum_a{T^{j*}_{am}T^{j}_{an}}, ~~\forall i, j, m, n, i\neq j,
\end{equation}
which says that the sum $\sum_a{T^{i*}_{am}T^{i}_{an}}$ does not depend on index i.

We can now combine conditions \ref{eq:condI} and \ref{eq:condII} into:
\begin{equation}
\sum_a{T^{i*}_{am}T^{j}_{an}}=\delta_{ij}C_{mn},
\end{equation}
and from the unitary constraints of Eq. \ref{eq:unitary} we see that $\sum_m{C_{mm}}=1$.

\emph{Step (c). Factorization:}

In this part, we will finally show the factorization of the wave-function.
First, we compute $\rho_C \equiv \tr_{AB} (\rho_{ABC})$,
\begin{eqnarray}
\notag
  \rho_C    &=& \sum_{ijka}{p_iT^i_{aj}T^{i*}_{ak}|\phi_C^j\rangle\langle \phi_C^k|} \\
\notag
            &=& \sum_{ijk}{p_iC_{jk}|\phi_C^j\rangle\langle \phi_C^k|} \\
            &=& \sum_{jk}{C_{jk}|\phi_C^j\rangle\langle \phi_C^k|}
\end{eqnarray}
Here we are free to choose the orthonormal basis $\ket{\phi_C^i}$ of part $C$ such that $\rho_C$ is diagonal, ie, the matrix $C_{ij}=\delta_{ij}q_j, \sum_i{q_i}=1$.

Let us now consider the set of states $|\tilde{\psi}_B^{ik}\rangle \equiv \sum_j{T^i_{jk}|\phi_B^j\rangle}$ of $\mathcal{H}_B^{\parallel}$. The scalar products
\begin{eqnarray}
\notag
  \langle \tilde{\psi}_B^{mn}|\tilde{\psi}_B^{ij}\rangle   &=& \left(\sum_a{T^{m*}_{an}\langle \phi_B^a|}\right) \left(\sum_b{T^i_{bj}|\phi_B^b\rangle}\right)  \\
\notag
                    &=& \sum_a{T^{m*}_{an}T^i_{aj}}  \\
\notag
                    &=& \delta_{mi}C_{nj}  \\
                    &=& \delta_{mi}\delta_{nj}q_j
\end{eqnarray}
reveal that they form an orthogonal basis in $\mathcal{H}_B^{\parallel}$, which we can normalize by defining $|\psi_B^{ik}\rangle \equiv |\tilde{\psi}_B^{ik}\rangle/\sqrt{q_k}$. The wave-function can now be written as
\begin{equation}
	\ket{\Psi_{ABC}} =\sum_{ij} \sqrt{p_iq_j}~ \ket{\phi_A^i ~ \psi_{B}^{ij} ~\phi_C^j}.
\end{equation}
Further, we can introduce a product basis $\ket{\phi_{B_1}^i} \otimes \ket{\phi_{B_2}^j} \equiv \ket{\psi_{B}^{ij}}$ which defines a factorization of $\mathcal{H}_B^{\parallel}$ into a tensor product $\mathcal{H}_{B_1} \otimes \mathcal{H}_{B_2}$. With respect to this decomposition, state $\ket{\Psi_{ABC}}$ factorizes as 
\begin{equation}
	\ket{\Psi_{ABC}} = \ket{\Psi_{A B_1}} \otimes \ket{\Psi_{B_2 C}}, \label{eq:fact2}
\end{equation}
where
\begin{eqnarray}
\ket{\Psi_{A B_1}} &\equiv& \sum_i \sqrt{p_i} \ket{\phi_A^i \phi_{B_1}^i},\\
\ket{\Psi_{B_2 C}} &\equiv& \sum_j \sqrt{q_j} \ket{\phi_{B_2}^j \phi_C^j }.
\end{eqnarray}
This completes the proof.

\hfill$\Box$

We end this section with two simple corollaries.

\textbf{\emph{Corollary 4}:} If a tri-partite pure state $\ket{\Psi_{ABC}}$ is such that $\mathcal{N}^{A|BC} = \mathcal{N}^{A|B}$, then $\mathcal{N}^{A|C} = 0$.

This can be seen from Eq. \ref{eq:fact2}, which implies that the density matrix $\rho_{AC} \equiv \tr_B (\rho_{ABC})$ decomposes as the product $\rho_{AC} = \rho_A \otimes \rho_C$.

The second corollary, below, can be proved by applying the disentangling theorem iteratively. Let  $|\Psi_{A_1A_2...A_n}\rangle \in \mathcal{H}_{A_1} \otimes \mathcal{H}_{A_2} \otimes \cdots \otimes \mathcal{H}_{A_n}$ be a pure state of a system that decomposes into $n$ parts $A_1$, $A_2$, $\cdots$, $A_n$. Let $\mathcal{N}^{A_1...A_i|A_{i+1}...A_n}$ denote the negativity between parts $A_1\cdots A_i$ and $A_{i+1} \cdots A_n$; and let $\mathcal{N}^{A_1...A_i|A_{i+1}}$ denote the negativity between parts $A_1\cdots A_i$ and $A_{i+1}$.

\textbf{\emph{Corollary 5}:}
The above entanglement negativities fulfill 
\begin{equation}
	\mathcal{N}^{A_1...A_i|A_{i+1}...A_n} = \mathcal{N}^{A_1...A_i|A_{i+1}}
\end{equation}
for all $i \in \{1,2, \cdots,n-1\}$ if, and only if, the state $|\Psi_{A_1A_2...A_n}\rangle$ factorizes as
\begin{equation}
|\Psi_{A_1A_2'}\rangle\otimes|\Psi_{A_2''A_3'}\rangle\otimes|\Psi_{A_3''A_4'}\rangle\otimes...
\otimes|\Psi_{A_{n-1}A_n}\rangle,
\end{equation}
where for each $i\in \{2,\cdots,n-1\}$, the vector space $\mathcal{H}_{A_i}$ decomposes as 
\begin{equation}
	\mathcal{H}_{A_i} \cong (H_{A_i'} \otimes H_{A_i''})\otimes H_{A_i}^{\perp}.
\end{equation}

\section{ Monogamy for Entanglement Negativity}

Corollary 4 implies that, in the specific context of the tripartite pure states addressed by the disentangling theorem, the entanglement negativity (somewhat trivially) satisfies the relation,
\begin{equation}
	(\mathcal{N}^{A|BC})^2 \geq (\mathcal{N}^{A|B})^2 + (\mathcal{N}^{A|C})^2. \label{eq:monogamy2b}
\end{equation}
for the particular case $\mathcal{N}^{A|C} = 0$, which saturates the inequality. If it was correct, then this inequality would tell us the following: when measuring entanglement by means of the squared negativity $\mathcal{N}^2$, if part $A$ is very entangled with part $B$, then part $A$ cannot be at the same time very entangled with part $C$. 

Eq. \ref{eq:monogamy} is reminiscent of (and motivated by) the famous Coffman-Kundu-Wootters monogamy inequality \cite{CKW} for another measure of entanglement, the concurrence $\mathcal{C}$ \cite{concurrence}, which satisfies 
\begin{equation}
 (\mathcal{C}_{A(BC)})^2 \geq (\mathcal{C}_{AB})^2+(\mathcal{C}_{AC})^2. \label{eq:C}
\end{equation}
According to this inequality, if $\mathcal{C}^2_{AB}$ approaches $\mathcal{C}^2_{A(BC)}$, then $\mathcal{C}^2_{AC}$ is necessarily small, which is used to say that entanglement is monogamous: if $A$ is very entangled with $B$, then it cannot be simultaneously very entangled with $C$. However, the concurrence $\mathcal{C}$ is only defined on qubits, and therefore of rather limited use. Thus in this section we explore whether the entanglement negativity, which is defined for arbitrary systems, can be used as a replacement of Eq. \ref{eq:C} for general systems. We first note that for the simplest possible system, made of three qubits, Y. C. Ou and H. Fan\cite{OuFan} showed analytically that Eq. \ref{eq:monogamy2b} holds. 

\begin{figure}
  \includegraphics[width=0.5\textwidth]{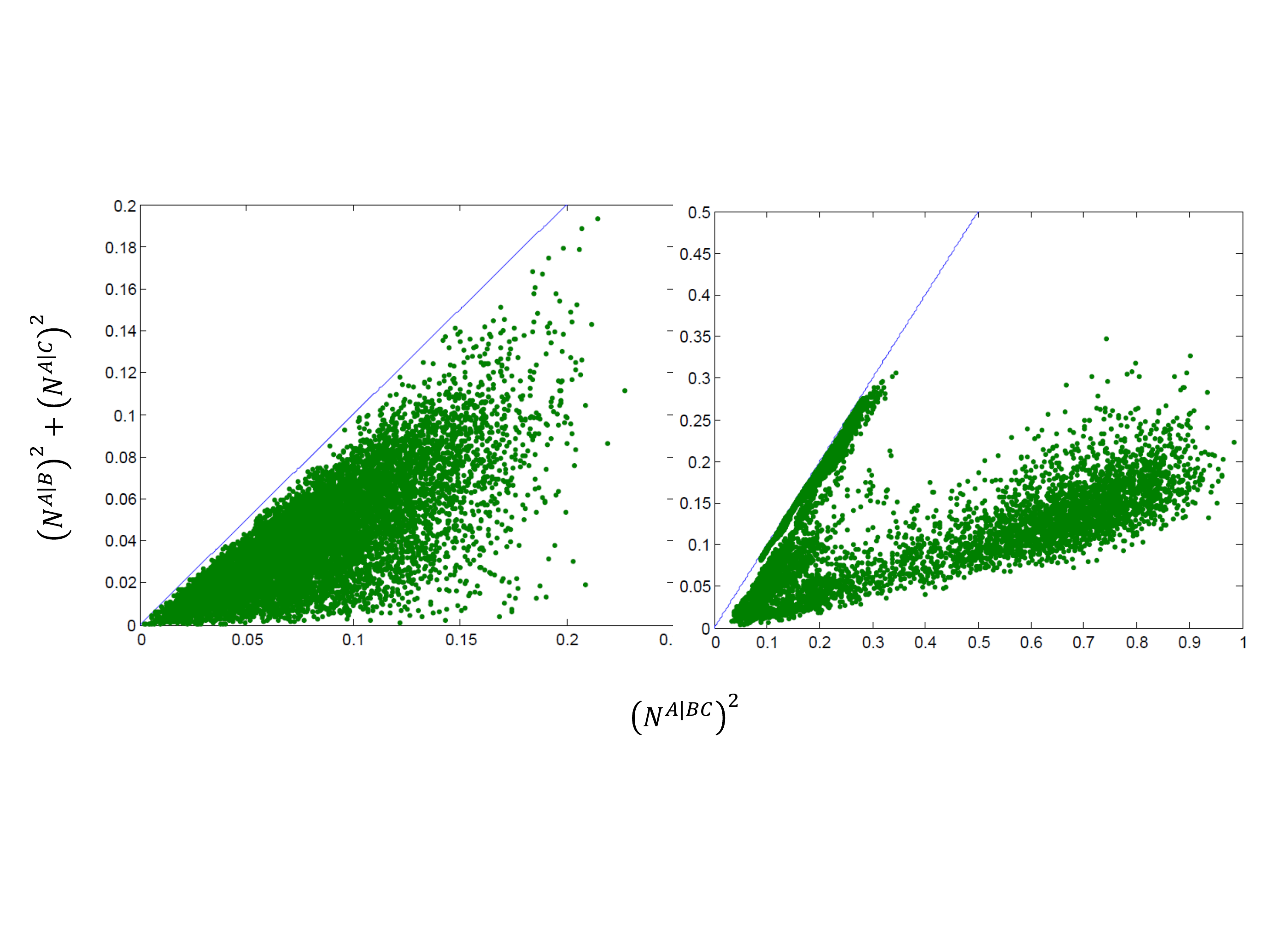}\\
  \caption{$(\mathcal{N}^{A|B})^2+(\mathcal{N}^{A|C})^2$ (y-axis) versus $(\mathcal{N}^{A|BC})^2$ (x-axis) for randomly generated pure states of three qubits $m=2$ (left) and three qutrits $m=3$ (right). The saturation line is $(\mathcal{N}^{A|BC})^2=(\mathcal{N}^{A|B})^2+(\mathcal{N}^{A|C})^2$ is represented in blue, while each green dot corresponds to a randomly generated pure states. These results are consistent with the general validity of Eq. \ref{eq:monogamy2b}. The concentration of points near the saturation line in the three-qutrit case is due to the sampling algorithm employed, which favors exploring the neighborhood of this saturation line.}
  \label{fig:monogamy}
\end{figure}

Here we address the validity of Eq. \ref{eq:monogamy2b} numerically. For $m=2,3$ and $4$, we have randomly generated hundreds of states in $\mathbb{C}_m \otimes \mathbb{C}_m \otimes \mathbb{C}_m$ and computed both sides of Eq. \ref{eq:monogamy2b}. The results for $m=2$ and $m=3$ are shown in Fig. \ref{fig:monogamy}. For $m=2$ our numerical results are consistent with the analytical proof presented in Ref. \cite{OuFan}. For $m=3$ we again see consistency with the monogamy relation \ref{eq:monogamy2b}, but a tendency already observed in $m=2$ becomes more accute: most randomly generated states concentrate away from the saturation line $(\mathcal{N}^{A|BC})^2 = (\mathcal{N}^{A|B})^2 + (\mathcal{N}^{A|C})^2$. This is a concern, because it means that we are not properly exploring the states near the saturation line, which are the ones that could violate the inequality. For this purpose, we used a Monte Carlo sampling whereby a new tripartite pure state similar to a previous one is accepted with certain probability that depends on the distance of its negativities to the saturation line.
The second panel in Fig. \ref{fig:monogamy} shows that this method indeed allows us to explore the neighborhood of the saturation line, and that there are again no violations of the monogamy relation \ref{eq:monogamy2b}. For $m=4$ (not displayed) similar results are obtained. We take these results as strong evidence of the validity of Eq. \ref{eq:monogamy2b} in the systems we have analyzed, and conjecture that Eq. \ref{eq:monogamy2b} should be valid for arbitrary tri-partite systems. Finally, we have also considered the generalized monogamy relation
\begin{equation}
	(\mathcal{N}^{A|B_1B_2 \cdots B_n})^2 \geq \sum_{i=1}^n (\mathcal{N}^{A|B_i})^2, \label{eq:monogamy3}
\end{equation}
in a system of $n+1$ parts $A$, $B_1$, $B_2$, $\cdots$, $B_n$. Specifically, we have numerically checked the validity of Eq. \ref{eq:monogamy3} for the case of four qubits ($n=3$).

\section{Conclusion}

In this paper we have provided two new results for the entanglement negativity. First, a disentangling theorem, Eq. \ref{eq:disentangling}, that allow us to use the negativity as a criterion to factorize a wave-function of a system made of three parts $A$, $B$, and $C$ into the product of two parts, namely $AB_1$ and $B_2C$, where we have also explained how to break the vector space of $B$ into those of $B_1$ and $B_2$. The second result is a conjectured monogamy relation, Eq. \ref{eq:monogamy2b}, which is known to hold for a system of three qubits \cite{OuFan} and that we have numerically confirmed for systems made of three $m$-level systems, for $m=2,3$ and 4.

These results are intended to add to our current understanding of entanglement negativity, at a time when this measure of entanglement is being consolidated as a useful tool to investigate and characterize many-body phenomena \cite{NegTN,NegMC}, including quantum criticality\cite{NegCFT} and topological order \cite{NegTO}.

\textbf{Acknowledgements}

Research at Perimeter Institute is supported by the Government of Canada through Industry Canada and by the Province of Ontario through the Ministry of Research and Innovation. G.V. acknowledges support from the Templeton Foundation. G.V. thanks the Australian Research Council Centre of Excellence for Engineered Quantum Systems.

\end{document}